\documentclass[10pt]{amsart}
\usepackage[cp866]{inputenc}
\usepackage[english,russian]{babel}
\usepackage{amsmath}
\usepackage{amssymb}
\usepackage{amsfonts}

\newtheorem{theorem}{Theorem}
\newtheorem{lemma}{Lemma}
\newtheorem{corollary}{Corollary}
\newtheorem{example}{Example}

\begin{document}
\renewcommand{\refname}{References}
\thispagestyle{empty}

\title{Canonical decomposition of catenation of factorial languages}
\author{{A. E. Frid}}%
\address{Anna E. Frid
\newline\hphantom{iii} Sobolev Institute of Mathematics,
\newline\hphantom{iii} pr. Koptyuga, 4,
\newline\hphantom{iii} 630090, Novosibirsk, Russia}%
\email{frid@math.nsc.ru}%

\thanks{\sc Frid, A. E.,
Catenation of factorial languages}
\thanks{\copyright \ 2006 Frid A. E.}
\thanks{\rm The work is supported by RFFI (grants 05-01-00364 and 06-01-00694).}

\maketitle {\small
\begin{quote}
\noindent{\sc Abstract. } According to a previous result by S. V.
Avgustinovich and the author, each factorial language admits a
unique canonical decomposition to a catenation of factorial
languages. In this paper, we analyze the appearance of the canonical
decomposition of a catenation of two factorial languages whose
canonical decompositions are given.
 \end{quote}
}

\section{Introduction}
This paper continues a research of decompositions of factorial
languages started in \cite{af,af2} and inspired by the field of
language equations and algebraic operations on languages in general (see, e.~g.,
\cite{kunc2,okh} and references therein). As the development of the theory
shows, even language expressions where the only used operation is catenation
prove very difficult to work with. It seems that nothing resembling the Makanin's
algorithm for word equations (see, e.~g., \cite{diekert}) can appear
for language equations with catenation. Even easiest questions tend
to have very complicated answers. In particular, the maximal solution $X$ of
the commutation equation
\[LX=XL\]
may be arbitrarily complicated: as it was shown by Kunc \cite{kunc},
even if the language $L$ is finite, the maximal language $X$ commuting with it
may be not recursively enumerable. This situation contrasts with
that for words, since $xy=yx$ for some words $x$ and $y$ implies
that $x=z^n$ and $y=z^m$ for some word $z$ and $n,m \geq 0$.

In some sense, the problems of catenation of languages are due to
the fact that a unique factorization theorem is not valid for it: as
it was shown by Salomaa and Yu \cite{s_yu}, even a finite unary language
can admit several essentially different decompositions to a
catenation of smaller languages, and an infinite language may have
no decomposition to {\it prime} languages and all; here a language $L$
is called {\it prime} if $L=L_1L_2$ implies that $L_1=\{\lambda\}$, where
$\lambda$ is the empty word, and $L_2=L$, or vice versa.

To avoid ambiguity of this kind, we restrict ourselves to
{\it factorial} languages. This family is large and widely
investigated since it includes, e.~g., languages of factors of finite or
infinite words and languages avoiding patterns (in the sense of
\cite{cas}). We can also consider the factorial closure of an arbitrary language.
Furthermore, the class of factorial languages is closed
under taking catenation, unit, and intersection, and constitutes a
monoid with respect to the catenation.

Decompositions of
factorial languages to a catenation of factorial languages
also may be several: for example, $a^*b^*=(a^*+b^*)b^*=a^*(a^*+b^*)$
(here and below $(+)$ denotes unit). However, as it was proved in
\cite{af}, we can define the notion of the {\it canonical}
decomposition of a factorial language which always
exists and is unique.

In this paper, we continue investigation of canonical decompositions
of factorial languages and solve the following general problem:
Given canonical decompositions of languages $A$ and $B$, what is the
canonical decomposition of their catenation $AB$?

Besides the self-dependent interest, the answer to this question may
help to solve equations on factorial languages. Indeed, equal
languages have equal canonical decompositions, and these canonical
decompositions may be compared as words. So, techniques valid for words
can be applied for them.

Thus, this paper may be considered as a description a tool helpful
for solving equations on factorial languages.

\section{Definitions and previous results}

Let $\Sigma$ be a finite alphabet, and $L \subseteq \Sigma^*$ be a
language on it. A word $u \in \Sigma^*$ is called a {\it factor} of
a word $v \in \Sigma^*$ if $v=sut$ for some (possibly empty) words
$s$ and $t$. The set of all factors of words of a language $L$ is
denoted by Fac$(L)$. Clearly, Fac$($Fac$(L))=$Fac$(L)$, so that
Fac$(L)$ may be called the {\it factorial closure} of $L$.

A language $L$ is called {\it factorial} if $L=$Fac$(L)$. In particular, each
factorial language contains the empty word denoted by $\lambda$.
In what
follows, we consider only factorial languages.

The catenation of languages is an associative operation defined by
\[XY=\{xy|x \in X, y \in Y\}.\]
Clearly, languages constitute a monoid with respect to the
catenation,  and its unit is the language $\{\lambda\}$, where
$\lambda$ is the empty word. It is also clear that factorial
languages form a submonoid of that monoid, since the catenation of two
factorial languages is factorial.

A factorial language $L$ is called {\it indecomposable} if $L=XY$
implies $L=X$ or $L=Y$ for all factorial languages $X$ and $Y$.

\begin{lemma}\cite{af}
For each subalphabet $\Delta \subseteq \Sigma$, the language
$\Delta^*$ is indecomposable.
\end{lemma}
Other examples of indecomposable languages discussed in \cite{af}
include languages of factors of recurrent infinite words, etc.

A decomposition $L=L_1 \cdots L_n$ to  factorial languages
$L_1,\ldots,L_n$ is called {\it minimal} if
\begin{itemize}
\item
$L=\{\lambda\}$ implies $n=1$ and $L_1=\{\lambda\}$;
\item
If $L \neq \{\lambda\}$, then for
$i=1,\ldots,n$ we have $L_i \neq \{\lambda\}$ and $L \neq L_1 \cdots
L_{i-1}L_i'L_{i+1}\cdots L_n$ for any factorial language $L_i'
\subsetneq L_i$.
\end{itemize}
A minimal decomposition to indecomposable factorial language is
called {\it canonical}.
\begin{theorem}\cite{af}
A canonical decomposition of each factorial language $L$ exists and
is unique.
\end{theorem}

In what follows, we shall denote the canonical decomposition of $L$ by
$\overline{L}$. Note that a canonical decomposition can be
considered as a word on the alphabet ${\mathcal F}$ of all
indecomposable factorial languages. In what follows, $(\doteq)$ will
denote equality of elements of ${\mathcal F}^*$; this notation will
be used to compare canonical
decompositions.

All examples of factorial languages we shall
consider in this paper will be regular, just because regular languages are easy to
deal with. Note that the factorial closure of a regular language is
always regular (which is a classical exercise). We have proved also

\begin{theorem}\cite{af2}
If $L$ is a regular factorial language, then all entries of $\overline{L}$ are also regular.
\end{theorem}

\section{Preliminary results}
Suppose that we are given two factorial languages, $A$ and $B$, on an alphabet
$\Sigma$, and know their canonical decompositions $\overline{A}$ and
$\overline{B}$. Our goal is to describe the canonical decomposition
$\overline{AB}$, and  the main result of the paper,
Theorem \ref{main}, will give such a description. To state Theorem \ref{main},
we need to define two
subalphabets of $\Sigma$, namely, $\Pi$ and $\Delta$.

For a factorial language $L$, let us define

\[\Pi(L)=\{a \in \Sigma|La \subseteq L\},\]
and
\[\Delta(L)=\{a \in \Sigma|aL \subseteq L\}.\]
Thus, if we take any word $u \in L$, we can extend it to the left by
any word from $\Delta^*(L)$ and to the right by any word from
$\Pi^*(L)$ to get a word from $L$. In other words,
$L=\Delta^*(L)L\Pi^*(L)$, and $\Pi(L)$ and $\Delta(L)$ are defined
as maximal languages with this property.

For the main result of this paper, we shall need to know the
relationship between $\Pi(A)$ (further denoted by $\Pi$) and
$\Delta(B)$ (further denoted by $\Delta$). The following lemmas
explain the meaning of these subalphabets. Note that analogues
of Lemmas \ref{aak}--\ref{l5}' were proved in \cite{af}, but the lemmas are
reproved here both for the sake of  completeness and of more precise wording.

\begin{lemma}\label{aak}
If $\overline{L}\doteq L_1\cdots L_k$, then
$\Pi(L)=\Pi(L_k)$ and $\Delta(L)=\Delta(L_1)$.
\end{lemma}
\noindent {\sc Proof.} Let us prove the statement for $\Pi(L)$; the statement
for $\Delta(L)$ is symmetric to it.

First, $\alpha \in \Pi(L_k)$ implies that $L_k \alpha \subseteq L_k$
and thus $L\alpha=L_1 \cdots L_k \alpha \subseteq L_1\cdots L_k =L$; so,
$\Pi(L_k) \subseteq \Pi(L)$.

On the other hand, $\alpha \in \Pi(L)$ means that $L_1\cdots L_{k-1} v\alpha
\subseteq L$ for all $v \in L_k$. Since $L_k$ is a factor of the
canonical decomposition of $L$, it cannot be contracted to a smaller
factorial language $L_k'$ such that $L_1\cdots L_{k-1} L_k' =L$. It
means that for each $v \in L_k \backslash \{\lambda\}$, there exists some
word $wtv \in L$ such that $w \in L_1\cdots L_{k-1}$, $tv \in L_k$,
and $w$ is the longest prefix of $wtv$ belonging to $L_1\cdots L_{k-1}$.
Since $tv$ is not the empty word, $w$ is also the longest prefix from
$L_1\cdots L_{k-1}$ of the
word $wtv\alpha \in L$. We see that $tv\alpha \in L_k$ and thus $v\alpha \in L_k$
since $L_k$ is factorial. Moreover, by the same reason $\alpha \in L_k$, which means
that $\lambda \alpha \in L_k$ and thus $L_k \alpha \subseteq L_k$. So, $\alpha \in
\Pi(L)$ implies $\alpha \in \Pi(L_k)$, which was to be proved. \hfill $\Box$

\medskip
Given a factorial language $A$ and a subalphabet $\Delta \subseteq \Sigma$, let
us define the factorial language $L_{\Delta}(A)=$Fac$(A\backslash \Delta
A)$. So, $L_{\Delta}(A)$ is the subset of $A$ containing exactly
words starting with letters from $\Sigma \backslash \Delta$ and
their factors.
Symmetrically, we define the
subset $R_{\Delta}(A)$ of $A$ containing exactly
words which end with letters from $\Sigma \backslash \Delta$ and
their factors: $R_{\Delta}(A)=$Fac$(A \backslash A\Delta)$.

\begin{lemma}\label{min}
Let $X$ and $B$ be factorial languages on $\Sigma$. If there exists
a factorial language $A$ such that $X=AB$, then there exists a
unique minimal one, and it is equal to $A'=R_{\Delta(B)}(A)$.
\end{lemma}
{\sc Proof.} First of all, let us prove that $A'B=X$. The
$\subseteq$ inclusion is obvious: $A' \subseteq A$ and thus $A'B\subseteq AB=X$.
To prove the $\supseteq$ inclusion, consider a word $x \in X$, and
let $b$ be its longest suffix from $B$: since $X=AB$, we have $x=ab$
for some word $a \in A$. Suppose that $a$ ends with a symbol $\delta
\in \Delta(B)$; then $\delta b \in B$ by the definition of
$\Delta(B)$, and $b$ is not the longest suffix of $X$ belonging to
$B$. A contradiction. Thus, $x=ab \in (A\backslash A\Delta(B))B \subseteq
R_{\Delta(B)}(A)B=A'B$, and since $x$ was an arbitrary element of
$X$, the $\supseteq$  inclusion (and thus the equality $X=A'B$) is proved.

It remains to prove that $A' \subseteq Y$ for every factorial
language $Y$ such that $YB=X$. Let us consider an arbitrary non-empty word $a'
\in A'$. Since $A'=R_{\Delta(B)}(A)$, the word $a'$ is a factor of
some word $sa't \in A \backslash A\Delta(B)$. Let the last letter of
the word $sa't$ be equal to $\alpha$; then $\alpha \in \Sigma
\backslash \Delta$, and $a't=a''\alpha \in A$. So, $a'tB \subseteq
AB=X=YB$.

For each $b \in B$, let us denote by $y(b)$ the longest prefix of
$a'tb=a''\alpha b$ belonging to $Y$. Let the word $b'$ be defined by the equality
$a'tb=y(b)b'$; then $b' \in B$ since $a'tb \in YB$.

Clearly, if $y(b)$ is not shorter than $a'$ for some
$b \in B$, then its prefix $a'$ belongs to $Y$ (since $Y$ is
factorial), and this is what we need. But if $y(b)$ is shorter than
$a'$ for all $b\in B$, then each word $b'$ contains $\alpha b$ as a suffix. So,
$\alpha b \in B$ for all $b \in B$
(since $B$ is factorial), and $\alpha \in \Delta(B)$ by the definition of
$\Delta(B)$. A contradiction. So, $a' \in Y$ for all $a' \in A'$,
and $A'$ is indeed the minimal language such that $A'B=X$. \hfill
$\Box$

\medskip
Symmetrically, we can prove

\noindent {\bf Lemma \ref{min}'} {\it
Let $X$ and $A$ be factorial languages on $\Sigma$. If there exists
a factorial language $B$ such that $X=AB$, then there exists a
unique minimal one, and it is equal to $B'=L_{\Pi(A)}(B)$}.

The following lemma is one of the main steps of the proof.
\begin{lemma}\label{eee}
For each factorial languages $A$ and $B$, we have
\[\overline{AB}\doteq \overline{R_{\Delta(B)}(A)}\cdot
\overline{L_{\Pi(R_{\Delta(B)}(A))}(B)}\doteq \overline{R_{\Delta(L_{\Pi(A)}(B))}(A)}\cdot
\overline{L_{\Pi(A)}(B)}.\]
\end{lemma}
{\sc Proof.} We shall prove the first equality; the second one can
be proved symmetrically.

Let us denote $R_{\Delta(B)}(A)=A'$ and
$L_{\Pi(R_{\Delta(B)}(A))}(B)=B''$. Due to Lemma \ref{min},
$A'B=AB$, and due to Lemma \ref{min}', $A'B''=A'B$. So, $AB=A'B''$.
Now note that all entries of the canonical decomposition of a language are
indecomposable. So, to prove the required equality of canonical decompositions
$\overline{AB}\doteq \overline{A'}\cdot \overline{B''}$, we must
prove only that no entry of the canonical decompositions
$\overline{A'}$ or $\overline{B''}$ can be decreased to get the same
product.

Indeed, suppose we substituted an indecomposable entry of
$\overline(A')$ by its proper factorial subset. Instead of $A'$, we obtained
its proper factorial subset $A_1$. Then $A_1 B \subseteq AB$ since $A'$ is the
minimal factorial language such that $A'B=AB$. But $B''\subseteq B$;
so, $A_1B''\subseteq A_1B \subsetneq AB$, and $A_1B''\neq AB$.

Now suppose we substituted an indecomposable entry of
$\overline{B''}$ by its proper factorial subset, and obtained a proper factorial
subset $B_1$ of $B''$. Then $A'B_1\neq A'B''=AB$ since $B''$ is the
minimal factorial set giving $AB$ when catenated with $A'$.

So, no entry of $\overline{A'}$ or $\overline{B''}$ can be replaced
by its proper subset without changing the result $AB$. The equality
is proved. \hfill $\Box$

\begin{lemma}\label{l5}
Let $X$ and $Y$ be factorial languages on $\Sigma$, and $\Delta \subset
\Sigma$ be a subalphabet such that $Y \nsubseteq \Delta^*$. Then
$R_{\Delta}(XY)=XR_{\Delta}(Y)$.
\end{lemma}
{\sc Proof.} Consider a word $u \in XR_{\Delta}(Y)$.
If $u \in X$, let us choose a symbol $y \in Y$ from $\Sigma \backslash
\Delta$. Then $uy \in XY\backslash XY\Delta\subseteq R_{\Delta}(XY)$, and
thus $u \in R_{\Delta}(XY)$. If $u \notin X$, then
$u=xu'$, where $x$ is the longest prefix of $u$ belonging to $X$ and
$u' \in R_{\Delta}(Y)$ is a non-empty word. Let $u''$ be
a word from $Y\backslash Y\Delta$ such that $u'$ is its factor:
$u''=su't$ for some words $s$ and $t$ such that the last letter of $t$
is from $\Sigma \backslash \Delta$. Then $u't \in Y\backslash
Y\Delta$, and hence $ut=xu't \in XY\backslash XY\Delta \subseteq R_{\Delta}(XY)$.
It follows that $u \in R_{\Delta}(XY)$, and the $\supseteq$
inclusion is proved.

To prove the $\subseteq$ inclusion, consider a word $u \in R_{\Delta}(XY)$.
Let $u'=sut$ be a word from $XY \backslash XY\Delta$
whose factor is $u$, so that its last letter is from $\Sigma \backslash
\Delta$. Then $ut \in XY \backslash XY\Delta$. Let $ut=xy$, where $x
\in X$ and $y \in Y$; then $y \in Y \backslash Y \Delta$ and $ut \in
X (Y \backslash Y \Delta)$. So, either $u \in X$, or $u=xy'$ for
some prefix $y'$ of $y$: since $y' \in R_{\Delta}(Y)$,
in both cases we have $u \in X R_{\Delta}(Y)$, and the
inclusion is proved. \hfill $\Box$

\medskip
Symmetrically, we prove

\noindent {\bf Lemma \ref{l5}'} {\it
Let $X$ and $Y$ be factorial languages on $\Sigma$, and $\Pi \subset
\Sigma$ be a subalphabet such that $X \nsubseteq \Pi^*$. Then
$L_{\Pi}(XY)=L_{\Pi}(X)Y$}.

The following series of lemmas is also one of important parts of the
main result.

\begin{lemma}\label{lxx}
Let $X$ be a factorial language, $\Pi \subset \Sigma$ be a subalphabet,
and $\Delta(X) \backslash \Pi \neq \emptyset$. Then $L_{\Pi}(X)=X$.
\end{lemma}
{\sc Proof.} Let $\alpha \in \Sigma$ be a
symbol from $\Delta(X) \backslash \Pi$; then each word $u$ from $X$ can
be extended to $\alpha u \in X$ by the definition of $\Delta(X)$.
So, $u \in $Fac$(\alpha u) \subset$Fac$(X\backslash \Pi X)=L_{\Pi}(X)$. Since
$u$ was chosen arbitrarily, and $L_{\Pi}(X) \subseteq X$, we get the equality:
$L_{\Pi}(X)=X$. \hfill $\Box$

\medskip
The symmetric lemma is

\noindent {\bf Lemma \ref{lxx}'} {\it
Let $X$ be a factorial language, $\Delta \subset \Sigma$ be a subalphabet,
and $\Pi(X) \backslash \Delta \neq \emptyset$. Then $R_{\Delta}(X)=X$.}

\begin{lemma}\label{case23}
For each factorial language $X$ with $\overline{X}\doteq X_1\cdots
X_k$ we have
\[\overline{L_{\Delta(X)}(X)}\doteq \left\{ \begin{array}{ll}
X_2\cdots X_k, & \mbox{~if~} X_1=\Delta^*(X),\\
\overline{X}, & \mbox{~otherwise.}
\end{array}\right.\]

Symmetrically,
\[\overline{R_{\Pi(X)}(X)}\doteq \left\{ \begin{array}{ll}
X_1\cdots X_{k-1}, & \mbox{~if~} X_k=\Pi^*(X),\\
\overline{X}, & \mbox{~otherwise.}
\end{array}\right.\]
\end{lemma}
{\sc Proof.} We shall prove the first equality; the second one is
symmetric. Let us denote $\Delta(X)=\Delta$.

Suppose first that $X_1\neq \Delta^*$, that is, $X_1\supsetneq
\Delta^*$. Due to Lemma \ref{l5}',
$L_{\Delta}(X)=L_{\Delta}(X_1)X_2\cdots X_k$. By the definitions,
$X_1=\Delta^*L_{\Delta}(X_1)$. But the language $X_1$ is
indecomposable and is not equal to $\Delta^*$, so,
$X_1=L_{\Delta}(X_1)$, and the equality $X=L_{\Delta}(X)$ (and thus
$\overline{L_{\Delta}(X)}\doteq \overline{X}$) is proved.

Now suppose that $X_1=\Delta^*$. Then $L_{\Delta}(X)=L_{\Delta}(X_2\cdots X_k)$
by the definition of the operator $L_{\Delta}$, since all elements of $X\backslash
X_2\cdots X_k$ cannot occur in $L_{\Delta}(X)$ anyway.
Then, $L_{\Delta}(X_2)=X_2$ because otherwise we would have
$X_1X_2=\Delta^*X_2=\Delta^*Y$ for some $Y=L_{\Delta}(X_2)\subsetneq
X_2$, contradicting to the minimality of the decomposition
$\overline{X}$. So, due to Lemma \ref{min}',
$L_{\Delta}(X)=L_{\Delta}(X_2\cdots X_k)=L_\Delta(X_2)X_3\cdots
X_k=X_2\cdots X_k$. The latter decomposition is minimal and thus canonical.
\hfill $\Box$

\section{Main result}
\begin{theorem}\label{main}
Let $A$ and $B$ be factorial languages with $\overline{A}\doteq A_1\cdots A_k$ and
$\overline{B} \doteq B_1\cdots B_m$. Let us denote $\Pi=\Pi(A)$ and
$\Delta=\Delta(B)$. Then the canonical decomposition of the
catenation $AB$ can be found as follows:
\begin{enumerate}
\item
If $\Delta \backslash \Pi \neq \emptyset$ and $\Pi \backslash \Delta \neq
\emptyset$, then $\overline{AB}\doteq \overline{A}\cdot \overline{B}$.
\item
If $\Delta=\Pi$, and $A_k\neq \Delta^*$, $B_1 \neq \Delta^*$, then
$\overline{AB}\doteq \overline{A} \cdot \overline{B}$.
\item
If $\Delta=\Pi$ and $A_k = \Delta^*$, then $\overline{AB}\doteq A_1\cdots A_{k-1}
\overline{B}$. Symmetrically, if $\Delta=\Pi$ and $B_1 = \Delta^*$, then
$\overline{AB}\doteq \overline{A}B_2\cdots B_m$.
\item
If $\Pi \subsetneq \Delta$, then $\overline{AB}\doteq
\overline{R_{\Delta}(A)} \cdot \overline{B}$.
Symmetrically, if $\Delta \subsetneq \Pi$, then
$\overline{AB}\doteq \overline{A}\cdot \overline{L_{\Pi}(B)}$.
\end{enumerate}
\end{theorem}
{\sc Proof.} Cases (1) and (4) are obtained directly by applying
Lemmas \ref{lxx} and \ref{lxx}' to the equality from Lemma
\ref{eee}. Case (2) is as well obtained by applying to Lemma
\ref{eee} Lemma \ref{case23}.

At last, in Case (3), if $A_k=\Delta^*$, we apply
Lemmas \ref{case23} and \ref{aak} to get
$\overline{L_{\Delta}(A)}\doteq A_1\cdots A_{k-1}$ and
$\Pi(L_{\Delta}(A))=\Pi(A_{k-1})$.
Assume that $\Pi(A_{k-1})$ includes $\Delta$ as a subset. Then
$A_{k-1}=A_{k-1}\Delta^*$, and $A=A_1\cdots
A_{k-1}\Delta^*=A_1\cdots A_{k-1}$, contradicting to the fact that
$\overline{A}\doteq A_1\cdots A_{k-1}\Delta^*$. So, $\Delta
\backslash \Pi(A_{k-1})\neq \emptyset$, and we apply Lemma \ref{lxx}
to get $L_{\Pi(A_{k-1})}(B)=B$. It remains to use Lemma \ref{eee} to
get Case (3) of the Theorem.\hfill $\Box$

\begin{corollary}\label{aorb}
The canonical decomposition of $AB$ either begins with $\overline{A}$,
or ends with $\overline{B}$, so that only one of the
languages $A$ and $B$ can give canonical factors of $AB$ different from the
canonical factors of the language itself.
\end{corollary}

\begin{example}
{\rm
If $A=\{a,b\}^*$ and $B=\{a,c\}^*$, then $\Pi(A)=\{a,b\}$,
$\Delta(B)=\{a,c\}$, and the canonical decomposition of $AB$ is just
$\{a,b\}^* \cdot \{a,c\}^*$ (Case (1)).
}
\end{example}
\begin{example}{\rm
If $A=$Fac$\{a,ab\}^*$ and $B=$Fac$\{a,ac\}^*$, then $\Pi(A)=\Delta(B)=\{a\}$,
and the canonical decomposition of $AB$ is just Fac$\{a,ab\}^*$Fac$\{a,ac\}^*$
(Case (2)).

Here $A$ is the language of all words on $\{a,b\}$ which do not
contain two successive $b$s, and $B$ is the language of all words on
$\{a,c\}$ which do not contain two successive $c$s.
}
\end{example}
\begin{example}
{\rm
If $A=a^*$ and $B=$Fac$\{a,ab\}^*$, then $\Pi=\Delta=\{a\}$, and
$AB=B$ (Case (3)).}
\end{example}
\begin{example}
{\rm
Note that when $\Delta=\Pi$ and $A_k=B_1=\Delta^*$, Case (3) may be applied in any of the
two directions. For example,
if $A=a^*b^*$ and $B=b^*a^*$, then $\overline{AB}\doteq a^*\cdot b^*\cdot a^*$, and it does not
matter which of the occurrences of $b^*$ was removed.
}
\end{example}

Before giving examples for Case (4), we will specify the form of the
canonical decomposition of $A'=R_{\Delta}(A)$. Recall that $A$ is a
factorial language with the canonical decomposition $\overline{A}\doteq A_1\cdots
A_k$, and $\Delta$ is a subalphabet of $\Sigma$.

Let us define languages
$A_i'$, $i=k,\ldots,1$, as obtained by the following
iterative procedure: starting from $\Delta_k := \Delta$, we put
for each $i$ from $k$ to 1
\begin{eqnarray*}
A_i'&=&R_{\Delta_i}(A_i) \mbox{~and~}
\Delta_{i-1}=\Delta(A_i'), \mbox{~if~} A_i \nsubseteq \Delta_i^*,\\
A_i'&=&\{\lambda\} \mbox{~and~} \Delta_{i-1} = \Delta_i,
\mbox{~otherwise}.
\end{eqnarray*}

\begin{lemma}\label{ca'}
The canonical decomposition of $A'=R_{\Delta}(A)$
can be obtained by deleting extra $\{\lambda\}$ entries
from the decomposition $A'=\overline{A_1'}\cdot \overline{A_2'} \cdots
\overline{A_k'}$.
\end{lemma}
{\sc Proof.} First of all, note that due to Lemma \ref{l5} applied
iteratively, $A'=A_1 \cdots A_{k-1} A_k' = A_1 \cdots A_{k-2}
A_{k-1}' A_k' = \ldots = A_1' \cdots A_k'$. Some of the languages
$A_i'$ can be equal to $\{\lambda\}$; in particular, if $A \subseteq
\Delta^*$, then $A'=\{\lambda\}$, as well as all its factors. However, if $A' \neq
\{\lambda\}$, then we can canonically decompose factors $A_i'$ not equal to $\{\lambda\}$
and erase the others.

Clearly, if we substitute any of canonical factors of $A_i'$ by its
proper subset, we get a new language $A_i'' \subsetneq A_i'$. So, to
prove the lemma, we should just show that $A' \neq A_1' \cdots
A_{i-1}' A_i'' A_{i+1}' \cdots A_k'$ for any $A_i'' \subsetneq
A_i'$.

For all $i=1,\ldots,k$, let us define $D_i=A_1' \cdots A_i'$ and
$E_{i-1}=A_1 \cdots A_{i-1}$. We
also define $D_0=\{\lambda\}$.
Note that by the definition and Lemma \ref{min}, for all $i\geq 1$,
$D_i$ is the minimal language such that $D_iA_{i+1}'
\cdots  A_k'=A'$. So, it remains to prove only that $A_i'=A_i''$, where $A_i''$
 is the minimal language such that $D_{i-1} A_i'' = D_i$. By Lemma
\ref{min}', we have $A_i''=L_{\Pi(D_i)}(A_i')$.

First, suppose that $D_{i-1} \neq E_{i-1}$. We knew that
$D_i=D_{i-1}A_i'=E_{i-1}A_i'$, and $D_{i-1}$ is the minimal language
giving $D_i$ when catenated with $A_i'$. So, by Corollary \ref{aorb},
in the canonical decomposition of $D_i$ the factors corresponding to
$\overline{A_i'}$ do not change, and $A_i'=A_i''$, which was to be proved.

Now suppose that $D_{i-1} = E_{i-1}$. Then
$\Pi(D_{i-1})=\Pi(E_{i-1})=\Pi(A_{i-1})$.
From now on, we denote this subalphabet just by $\Pi'$.
We knew that $A_i$ was equal to $L_{\Pi'}(A_i)$ since
it was the minimal factorial language giving $E_i$ when catenated
with $E_{i-1}$. Assume by contrary that
$A_i''=L_{\Pi'}(A_i') \neq A_i'$.

Let us consider a word $u \in A_i' \backslash A_i''$. It does
not belong to $A_i''$, which means that $su \in A_i'$ implies $su \in \Pi' \Sigma^*$ for all
$s \in \Sigma^*$ (in particular, $u$ starts with a letter from $\Pi'$).
On the other hand, $u \in A_i'$, which means that
$ut \in A_i \cap A_i (\Sigma \backslash \Delta_i)$ for some $t \in
\Sigma^*$. By the definition, $ut \in A_i'$, and the set of non-empty left
extensions of $ut$ to elements of $A_i$ is a subset of that for $u$:
\[\{s \in \Sigma^+|sut \in A_i\} \subseteq \{s\in \Sigma^+|su \in
A_i'\} \subseteq \Pi' \Sigma^*.\]
Since we already know that $\lambda u =u \in \Pi' \Sigma^*$, we see
that $ut \notin L_{\Pi'}(A_i)$. So, $A_i \neq L_{\Pi'}(A_i)$,
contradicting to the fact that the decomposition $E_i=A_1
\cdots A_i$ was minimal. We have found a contradiction to the assumption that
$A_i' \neq A_i''$.

So, $A_i'=A_i''$, and the decomposition obtained from $A'= A_1'\cdots A_k'$ by
deleting $\{\lambda\}$ entries is minimal, which was to be proved. \hfill $\Box$

To make the description complete, we state the symmetric lemma, for
the case of $\Delta \subsetneq \Pi$. Let $B$  be a factorial
language with $\overline{B}\doteq B_1\cdots B_m$
and $\Pi$ be a subalphabet; we start from $\Pi_1=\Pi$ and successively define
for each $j=1,\ldots,m$
\begin{eqnarray*}
B_j'&=&L_{\Pi_j}(B_j) \mbox{~and~}
\Pi_{j+1}=\Pi(B_j'), \mbox{~if~} B_j \nsubseteq \Pi_j^*,\\
B_j'&=&\{\lambda\} \mbox{~and~} \Pi_{j+1} = \Pi_j,
\mbox{~otherwise}.
\end{eqnarray*}
The lemma symmetric to Lemma \ref{ca'} is

\noindent {\bf Lemma \ref{ca'}'} {\it
The canonical decomposition of $B'=L_{\Pi}(B)$
can be obtained by deleting $\{\lambda\}$ entries
from the decomposition $B'=\overline{B_1'}\cdot \overline{B_2'}\cdots
\overline{B_m'}$.
}

\medskip
The following easy example for Case (4) of Theorem \ref{main} illustrates Lemma
\ref{ca'}.

\begin{example}{\rm
The canonical decomposition of $A= (a^*b^*)^k+(b^*a^*)^k$ is
$\overline{A} \doteq (a^*+b^*)^{2k}$ with $A_1=\cdots=A_{2k}=(a^*+b^*)$ (here $+$
denotes the unit). If we catenate it with $B=a^*$, we get
$A_{2k}'=b^*$, $A_{2k-1}'=a^*$, and so on, and at last obtain
$A'=(a^*b^*)^k$ and $\overline{AB}\doteq (a^*\cdot b^*)^k \cdot a^*$.
}
\end{example}

\end{document}